\documentclass[11pt]{eptcs}
\usepackage{mathtools}
\usepackage{amssymb}
\usepackage{graphicx}
\usepackage{amsmath}
\usepackage{comment}
\usepackage{appendix}
\usepackage{subfigure}
\usepackage{float}
\usepackage{natbib}
\usepackage[font={small}]{caption}
\bibpunct[, ]{[}{]}{;}{n}{,}{,}

\title{Formal languages analysed by quantum walks}

\begin{document}

\begin{center}
\textbf{{\Large Formal languages analysed by quantum walks}}\ 
\
\vspace{5mm}

K. Barr    \hspace{10mm}     V. Kendon\
\vspace{3mm}

{ \small School of Physics and Astronomy, E. C. Stoner Building, University of Leeds, Leeds 
LS2 9 JT}\
\vspace{3mm}

\end{center}

\vspace{3mm}
\begin{center}
\textbf{{ \large Abstract }}

\end{center}

Discrete time quantum walks are known to be universal for quantum computation. This has been proven by showing that they can simulate a universal gate set. In this paper we examine computation in terms of language acceptance and present two ways in which discrete time quantum walks can accept some languages with certainty. These walks can take quantum as well as classical inputs, and we show that when the input is quantum, the walks can be interpreted as performing state discrimination.

\section{Introduction}

Both continuous and discrete time quantum walks are known to be Turing universal \cite{Lovett10, Childs09, Childs12}. This has been proven by showing that in both cases the elementary universal gate set can be simulated by a quantum walk. This maps the quantum walk onto the quantum circuit model. Both walks propagate amplitude deterministically along wires punctuated by gates formed using an appropriate combination of graph structure and coin operation. Quantum circuits have, despite much effort, not yet been realised experimentally beyond a few qubits. The best quantum computation devices currently in existence are based on liquid state NMR \cite{NMR}. Other models of computation better model these devices, in particular the Latvian Quantum Finite Automaton (LQFA) \cite{Algebraic}, which has been specially designed for that purpose. LQFAs are characterised in terms of language acceptance.

Due to the construction in \cite{Lovett10} we know that there must exist a mapping from a quantum walk onto any other Turing universal model of computation. In this paper we discuss work in progress concerning discrete time quantum walks in relation to language acceptance problems. We show that there are multiple ways to do this by exploring a range of small examples and that exploiting explicitly quantum aspects of the walks allows for gains in efficiency. We then introduce the concept of quantum inputs, and describe a preliminary investigation into the effect these have on word acceptance.

The paper is arranged as follows: First we briefly introduce formal languages and discrete time quantum walks. Then we give examples of languages which can be accepted by quantum walks in two formats and discuss their efficiency. We note that the constructions enable quantum inputs, and briefly indicate how these might be used to apply the walks to state discrimination \cite{Chefles00, QWPOVM}. The paper concludes with a summary and outline of future work.

\section{Formal languages}

The simplest type of formal languages are regular languages. Regular languages from alphabet $\Sigma$ are denoted by \textit{regular expressions}:

\begin{itemize}
\item{$\emptyset$: the empty set}
\item{$\epsilon$: the empty symbol}
\item{$a \in \Sigma$: singleton symbol}
\end{itemize}

From regular expressions $a$ and $b$ further regular expressions are formed inductively. If $a$ and $b$ denote regular expressions, their concatenation $a + b$, their union $a \cup b $  and their Kleene closure $ \{a, b \}^* $ are regular expressions. For example, the regular expression $\{ab\}^*$ denotes the language $\mathcal{L}_{ab} = \{ (ab)^m | m \in \mathbb{N} \}$. Languages can contain an arbitrary number of symbols but we mostly limit our considerations to languages expressible in binary. 

Regular languages are recognised by the simplest form of theoretical computer, the finite state automaton (FSA). FSAs have quantum counterparts known as quantum finite automata, of which there are many varieties \cite{Algebraic, BP, DQC, KW, WOMpaper}. Not all of these varieties are as strong as their classical analogue, but some such as the QFA-WOM \cite{WOMpaper} are stronger, accepting the context-free language $ \mathcal{L}_{eq} =  \{a^m b^m | m \geq 1 \}$ discussed in Sections \ref{sec:spatial}  and \ref{sec:sequ} below. Context free languages require an automaton with a `first in/last out' memory, known as a pushdown automaton, in order to be accepted. Universal Turing machines accept any language that is recursively enumerable.

FSAs end in a specific state, so they either fully accept an input, or they do not. The case with probabilistic outputs, such as in probabilistic finite automata (PFA) is a little more complex. The language $\mathcal{L}$ recognised with cut-point $\lambda \in [0, 1)$ by the PFA $M$ is $\mathcal{L}=\{ w|w \in\Sigma^* \, f_M(w)> \lambda \}$ where $f_{M}$ is the probability of acceptance. $M$ is said to accept $\mathcal{L}$ with bounded error if there is some $\epsilon> 0$ such that for all $w \in\mathcal{L}$, $M$ accepts $w$ with probability greater than $\lambda + \epsilon$ and accepts all words $w 􏰗 \notin \mathcal{L}$ with probability less than $\lambda - \epsilon$ where $\epsilon$ is the error margin. If there is no such $\epsilon$ then $M$ accepts with unbounded error.


\section{Discrete time quantum walks}
\label{sec:discqw}

For an arbitrary graph $G= \{ E, V \}$ a quantum walk evolves according to local operator $U= SC$. The coin operator $C$ is any unitary of dimension $d$ where $d$ is the degree of the node.  A different coin can be applied at every node. The shift operator permutes the position states according to $S|a, v \rangle = |a, u \rangle$ if $u$ is the $a^{th}$ neighbour of $v$ \cite{Gengraphs}.

To find a discrete time quantum walk which can differentiate strings of a given type, suitable definitions of input and output must be specified. Then a graph structure and coin operations which transform the input into an appropriate output must be found. There are a variety of ways that this can be done. For example, using a single walker starting in a superposition the input can be distributed along different nodes so that the entire word is operated on simultaneously; or the input symbols can be fed into the graph structure one after another. These could potentially be implemented using a multiport interferometer, modulated to enable differentiation between $a$ and $b$ inputs. 

The simplest way to deal with the output appears to be to designate an `accepting node,' which all accepting amplitude should be directed to. The problem then is to find graph structures and sets of coin operations which, upon a string from the language the walk is designed to accept, transports a high proportion of amplitude to the accepting node. If the remaining amplitude can be redirected to another, rejecting, node then the accepting and rejecting conditions can be inverted to allow the same walk to accept both its language and its complement- the set of words not in that language. Choosing these methods of input and acceptance turns the language acceptance problem into a perfect state transfer problem, see, for example \cite{ PSTrev, Bose03, Christandl05}.


\newpage
\section{Spatially distributed input}
\label{sec:spatial}
\vspace{-6mm}
\begin{center}
\begin{figure}[t]
\subfigure a) \includegraphics[scale = 0.25]{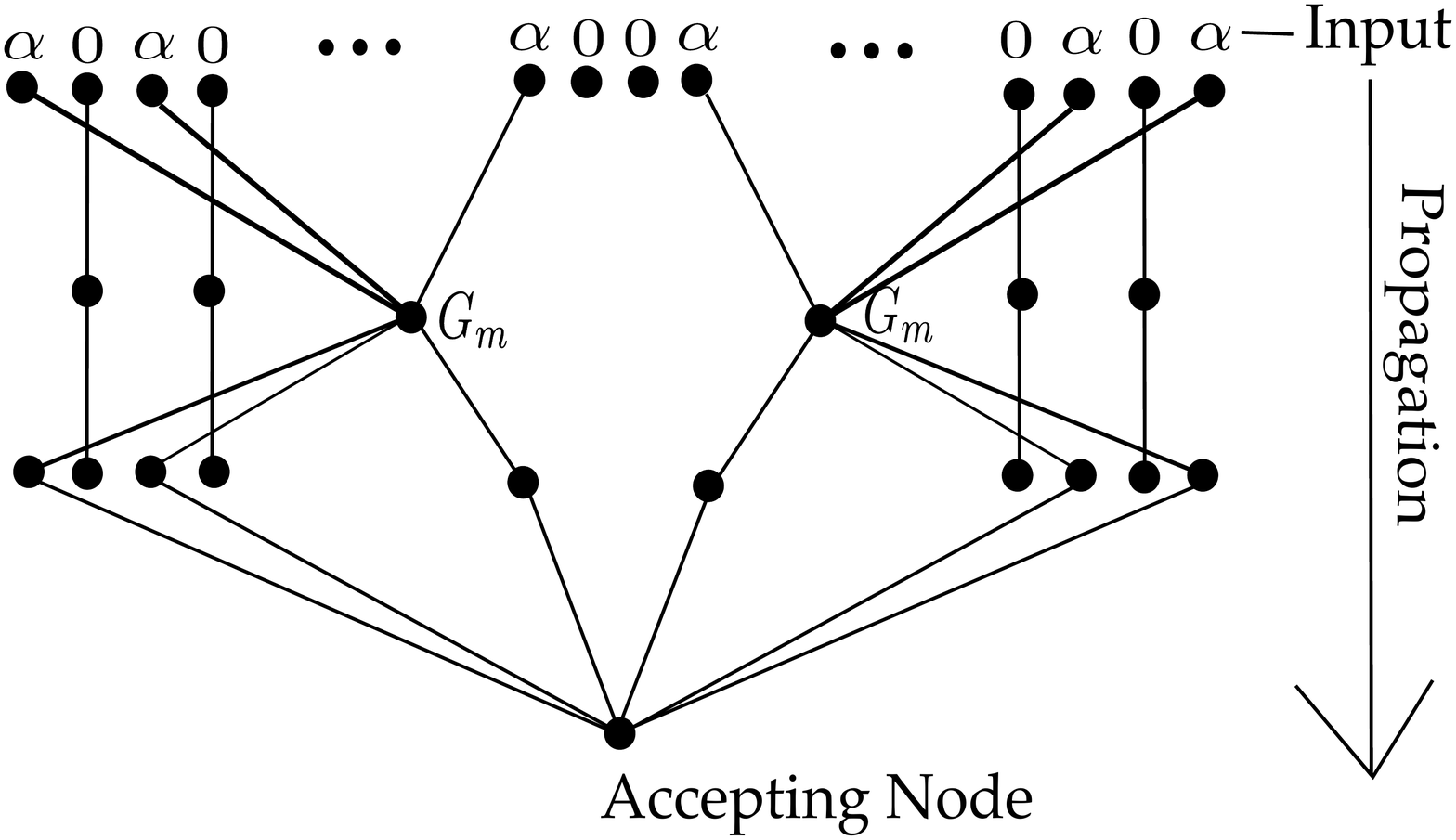} \subfigure b) \includegraphics[scale=0.25]{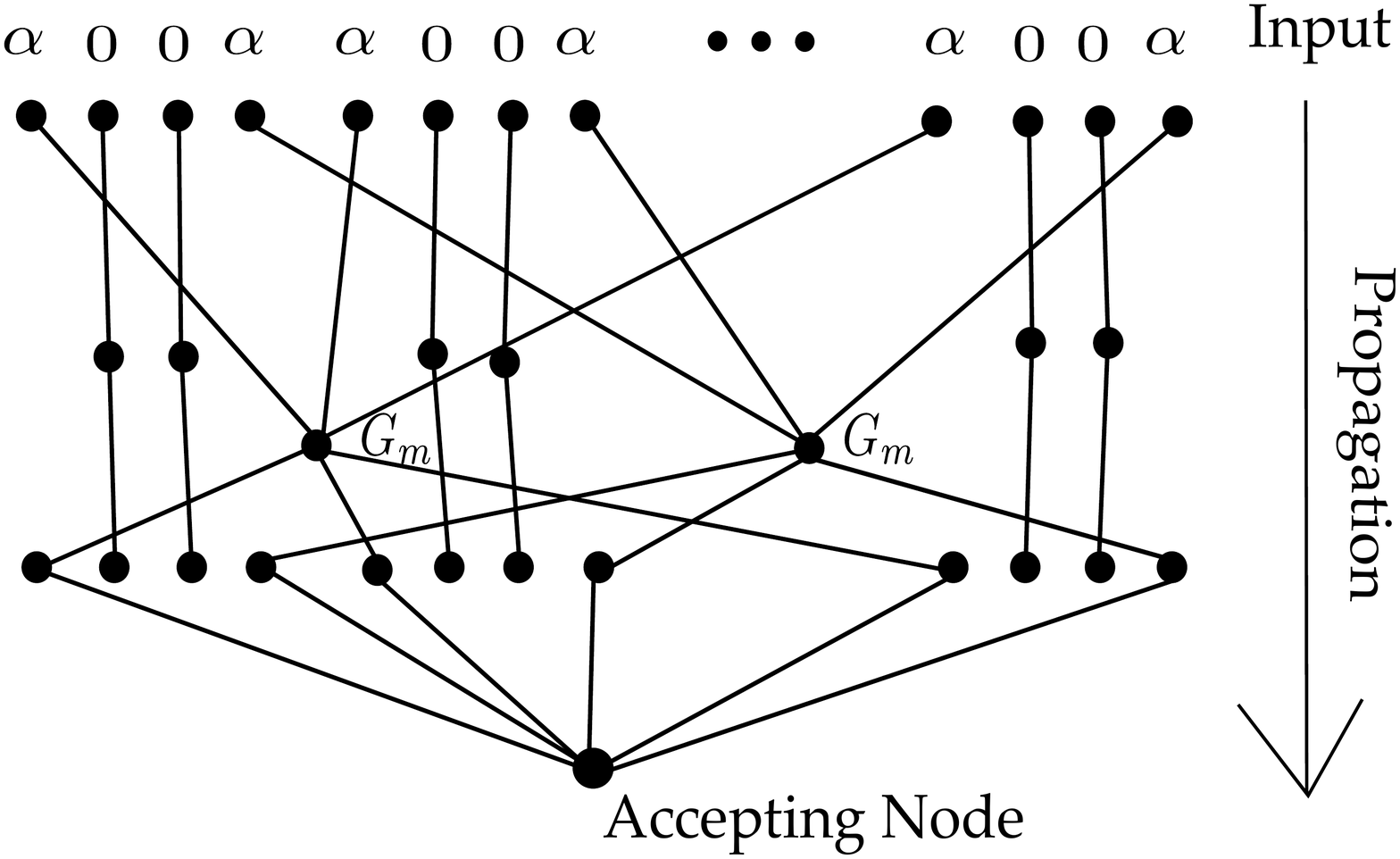}
\caption{Graphs required to accept a) $\mathcal{L}_{eq}$ and b) $\mathcal{L}_{ab}$ with each input symbol being operated on simultaneously. Grover coins of the appropriate dimensions are used at each vertex. Amplitude encodes a word which will be accepted by the walk. Edges join only at nodes indicated by black circles} \label{fig:ambmseq}
\end{figure}
\end{center}\

The input can be initially distributed along different nodes of the graph which are all operated on simultaneously. There are two nodes for each input symbol with alternate nodes representing $a$ or $b$. The length $n$ of the input must be known. For each input symbol $j$, the node $2j$ is populated with amplitude $1/\sqrt{n}$ if $j$ is an $a$ and $2j + 1$ contains amplitude if it's a $b$. The structure of the graph and coin for a walk testing whether a word is in a given language must be specifiable in terms of $n$ only. The walk is then run for a predetermined number of steps. When the initial state encodes a word in the language accepted by the walk, the amplitude should be directed to the designated `accepting node.' If a word not in the language accepted is encoded in the initial state then less of the amplitude will be directed towards the accepting node, so the acceptance is with bounded error. The modulus squared of the final amplitude at the accepting node yields the probability of acceptance.

These walks are most easily used to swiftly accept languages which contain at most one word of each length, so the graph tests for that specific word. Examples of such languages are the context free language $\mathcal{L}_{eq} = \{ a^m b^m | m \in \mathbb{N}\}$ and the regular language $ \mathcal{L}_{ab}= \{ (ab)^m | m \in \mathbb{N}  \} = \{ab\}^*$ and the graphs accepting them can be seen in Figure \ref{fig:ambmseq}. The input is accepted with certainty if it is in the language accepted by the walk. The probability of accepting words not in the language depends precisely on the string, but will for strings of length $n >1$ be less than or equal to $2/n^2$.

In both of these walks the input is processed in three full steps of the walk, regardless of length. However the gains in time complexity are at the expense of spatial complexity. For inputs of length $n$ the walks accepting $\mathcal{L}_{eq}$ and $\mathcal{L}_{ab}$ require $4n + 3$ nodes. The graph from Figure \ref{fig:ambmseq} a) can be extended to accept the archetypal context-sensitive language $\mathcal{L} = \{ a^m b^m c^m | m \in \mathbb{N} \}$ and b) to accept $\mathcal{L} = \{ (abc)^m | m \in \mathbb{N} \} = \{ abc \}^*$. In this case we must extend the model to deal with more than two input symbols which will involve a corresponding increase in the number of nodes required. 

The properties of the walks over the graphs indicated in Figure \ref{fig:ambmseq} were investigated by simulating them in Python for all possible inputs of a given length, so the algorithms are verified numerically. For both languages the results were very similar so we limit the discussion to $\mathcal{L}_{eq}$ here. The fidelity between the state obtained and the accepting state was calculated, where the fidelity between states $| \psi \rangle $ and $| \phi \rangle $ is defined as:

\begin{equation}
F =|  \langle \phi | \psi \rangle |^2
\label{eq: fid}
\end{equation}

The Jaro distance between the input and the closest word to a word in the language under consideration for that length was also calculated. The Jaro distance of strings $w_1$ and $w_2$:

\begin{equation}
d_j = \left\{ \begin{array}{cc} 0  & \text{if } m = 0 \\
\frac{1}{3} ( \frac{m}{|w_1|} + \frac{m}{|w_2|} + \frac{m-t}{m}   ) & \text{otherwise} \end{array} \right.
\end{equation}
with $m$ being the number of matching characters, characters which occur in both strings, in the same order, within a certain distance determined by the length of the strings. The value of $t$ is obtained by dividing the number of characters which differ by sequence order by 2. The three parts to the equation calculate the ratios of the number of matching characters to the lengths of $w_1$ and $w_2$ and then the ratio of non-transpositions to matching characters. 

The Jaro distance was selected as it always has values between 0 and 1, with 1 indicating that two words are equal, hence it was easy to compare to the fidelity. In the case of even length inputs, this simply was the word from $\mathcal{L}_{eq}$ of that length, for odd inputs with length $n$ the word was compared to the word in $\mathcal{L}_{eq}$ of length $n-1$. The results for the first 200 strings are illustrated in Figure \ref{fig:ambmspa} below. The points at which both curves peak are at the position of the words $ab$, $aabb$, $aaabbb$. 

The disparity between the Jaro distance and the fidelities for words not in $\mathcal{L}_{eq}$ arises from the design of the algorithm, which requires a low probability of acceptance for any word not in the language, regardless of how close that word is to a word in $\mathcal{L}_{eq}$. Hence the fidelity cannot be used as a good measure of how close the input word is to one in in that language in cases where it is not equal to 1. 

The constructions used here are based on the fact that a $d$ dimensional Grover operator:
\begin{center}
\begin{equation}
\label{eqn:grover}
G_{d}= \begin{pmatrix}
 \frac{2-d}{d} & \frac{2}{d} & \cdots & \frac{2}{d} \\
& & &\\
  \frac{2}{d} & \frac{2-d}{d} & \cdots & \frac{2}{d} \\
& & & \\
  \vdots  & \vdots  & \ddots & \vdots  \\
& & & \\
   \frac{2}{d}  & \frac{2}{d} & \cdots & \frac{2-d}{d}
 \end{pmatrix}
\end{equation}
\end{center}
 with $d$ even, transmits all amplitude to the `leaving' coin states of a vertex if the amplitude is initially evenly (with respect to both magnitude and phase) distributed between $n/2$ `entering' coin states. The deterministic evolution that the Grover operator can produce was used to design the `wires' which transmitted the amplitude between gates in \cite{Lovett10}. This fact can also be exploited to generate walks accepting with certainty other languages such as $\mathcal{L}_{twin} = \{ ww | w \in \{a, b\}^* \} $ and $\mathcal{L}_{rev} = \{ ww^r | w \in \{a, b\}^* \} $ where $w^r$ denotes the symbols of $w$ in reverse order. 

Using a spatially distributed input allows for long words to be accepted with the same number of operations as short words. However, the number of nodes required in the graph structure grows, albeit polynomially, with the length of the word. The number of nodes required to accept a given language can be held constant regardless of input length if each input symbol is fed into the structure in turn, so we now turn to this case.

\begin{center}
\begin{figure}[t]
\centering \includegraphics[scale = 0.4]{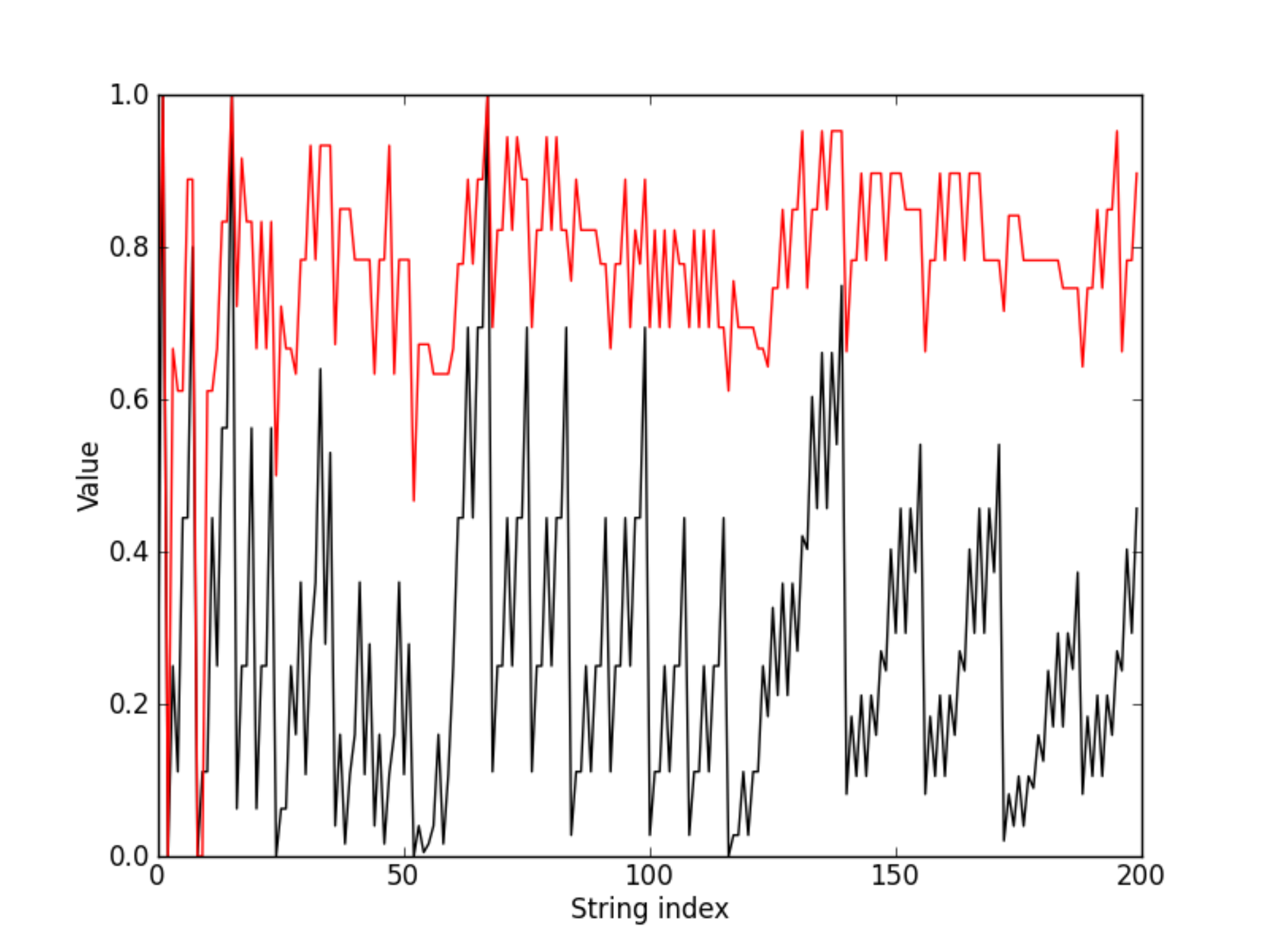}
\caption{Fidelity of final state to accepting state for walk detecting word $\mathcal{L}_{eq}$ for the first 200 strings (black). The Jaro distance between the input word and an appropriately sized word from $\mathcal{L}_{eq}$ is indicated in red. Both curves go to unity at the positions representing the strings $ab$, $aabb$ and $aaabbb$.} 
\label{fig:ambmspa}
\end{figure}
\end{center}


\section{Sequentially distributed input}
\label{sec:sequ}

The input can be treated sequentially if we start with it along a chain with two links between each node. The two symbols are represented:

\begin{equation} 
a = \begin{pmatrix} \alpha \\ 0  \\ 0 \\ 0 \end{pmatrix} \,\,\,\,\,\,\,\,\,\,\,\,\,\,\,\,\,\,\,  b = \begin{pmatrix} 0 \\ \alpha \\ 0 \\ 0 \end{pmatrix}  ~\label{eqn:abs}
\end{equation} \
\vspace{5mm}

The coin on this part of the graph is $ \sigma_x \otimes \mathbb{I}_2$, which simply swaps amplitude between the `leaving' nodes of the current node to the `arriving' nodes of the next node. These are then fed into a graph which has either  `accepting paths' and `rejecting paths' or 'accepting nodes' and `rejecting nodes.' The total square of the modulus of the amplitude on the accepting node/path gives the probability of accepting the input. The shape of the graph and the coins at each node determine which words will be accepted. 

%
%

\begin{center}
\begin{figure}[h] \subfigure a) \includegraphics[scale = 0.29]{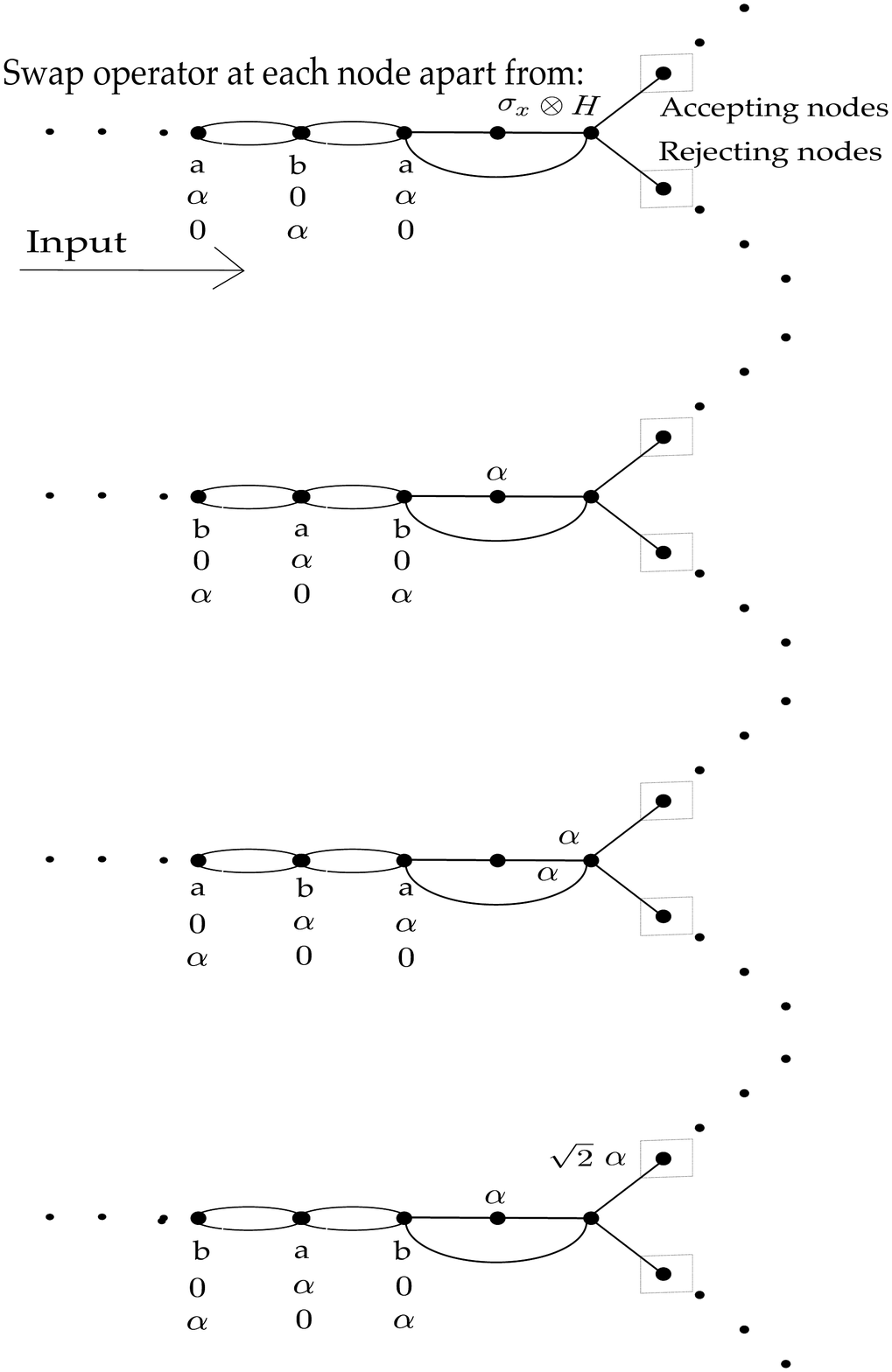} \subfigure b) \includegraphics[scale=0.29]{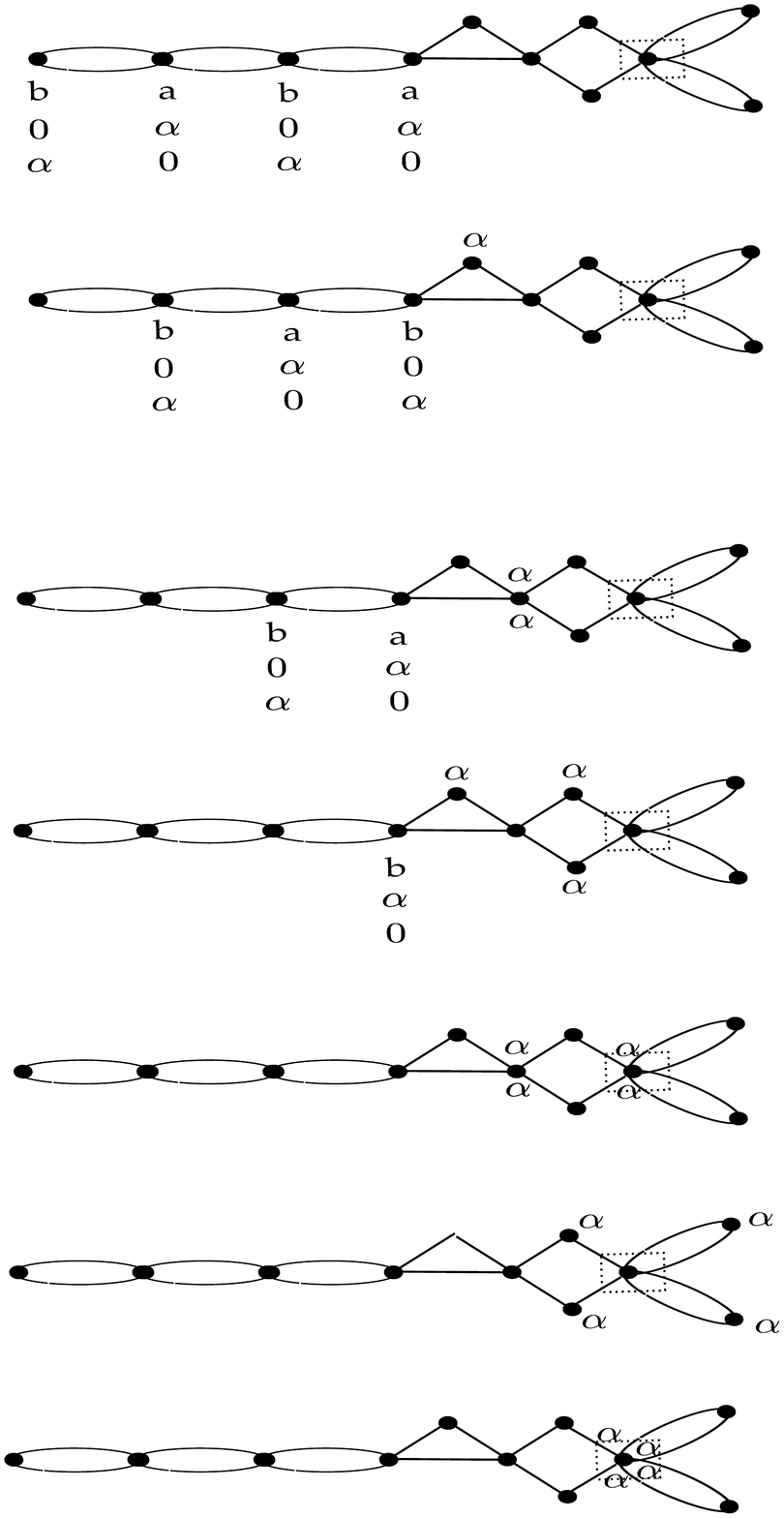}
\caption{a) Graph structure and coins accepting language $\mathcal{L}_{ab}$ b) Graph specifically accepting the word of length 4 from that language} 
\label{fig:abmstr}
\end{figure}
\end{center}

The empty set and empty string are accepted trivially, as we can distinguish between no walk occurring and all amplitude being rejected. Singleton symbols can be accepted by paths of length 3 with the amplitude initially in the appropriate coin state of the central node and swap operators between each node.  In some cases, such as walks accepting specific strings of known length, the only coins required are trivial swap operators, as in Figure \ref{fig:abmstr} (b). More complex languages require more complex graphs, such as that in Figure \ref{fig:abmstr} (a), the graph accepting the language $\mathcal{L}_{ab}$. This graph uses the Hadamard operator:

\begin{equation}
H = \begin{pmatrix}  \frac{1}{\sqrt{2}} & \,  \frac{1}{\sqrt{2}}  \\ \frac{1}{\sqrt{2}}  & \, \frac{-1}{\sqrt{2}} \\ \end{pmatrix}
\end{equation}
to determine whether each pair of symbols fed to it is of the form $ab$ and moves the amplitude from both symbols into the accepting path if they are of that form. Words in the language are accepted with certainty and those not in the language are accepted with probability $1/2$. By adding $m$ nodes rather than one to the path which amplitude from $a$'s goes into, this walk can be modified to accept $\mathcal{L}_{eq}$

Here we have an example of multiple graphs existing which accept a specific word. Either a specific graph or the graph accepting $ \mathcal{L}_{ab} $ can be used to accept $abab$, as shown in Figure \ref{fig:abmstr} (a) and (b) below. In the case tested for comparing whether different graphs accepting the same word give rise to different probability distributions, the probability of acceptance did not depend on the graph, however this property is not generally expected to hold. Here we can see that exploiting quantum properties, using the Hadamard operator to control interference between different parts of the amplitude rather than permuting amplitude so that it all arrives at the right place eventually, gains us efficiency. The simple swapping version of the graph accepting $abab$ has $8$ nodes (discounting the input nodes) and takes $6$ steps to accept the word. Whereas the more general graph not only accepts more words, but accepts $abab$ in $5$ steps. To accept longer words from $\mathcal{L}_{ab}$ using a permutation scheme rather than the graph using the Hadamard operator, more cycles are required and the size of the graph increases accordingly.

\begin{center}
\begin{figure}[t]
\centering \includegraphics[scale = 0.4]{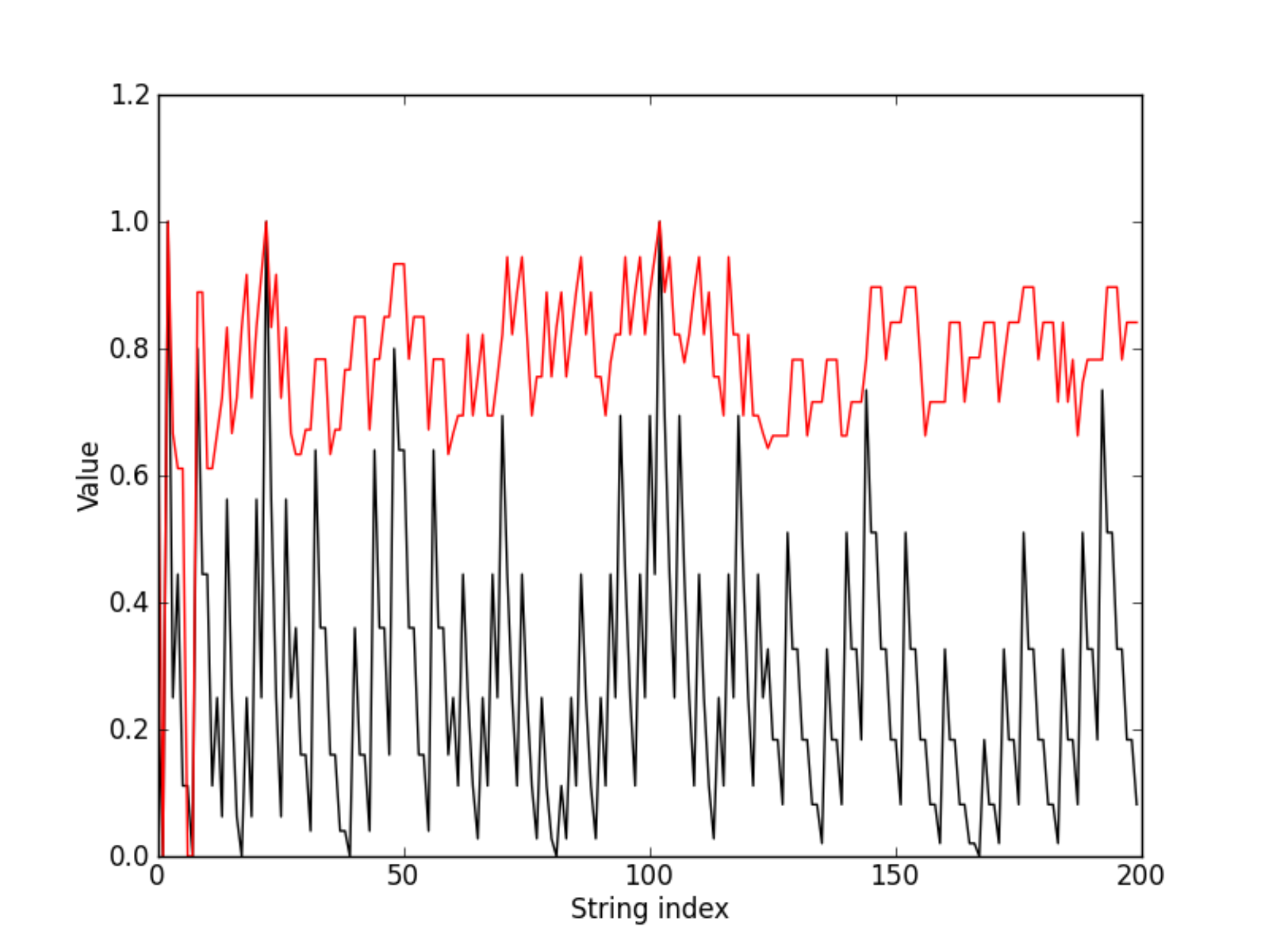}
\caption{Fidelity of final state to accepting state for walk detecting word from $\mathcal{L}_{ab}$ for the first 200 strings (black). The Jaro distance between the input word and an appropriately sized word from the language is indicated in red.} 
\label{fig:abmseq}
\end{figure}
\end{center}

We demonstrate the algorithms correctness enumeratively, but the result can be proven easily by induction on $m$, where $m$ is the number of times the string $ab$ is repeated. Again the fidelity between the final state for each input string and the required state for a string of that length was computed. This was plotted alongside the Jaro distance from the input word to a string of an appropriate length in the language as can be seen in Figure \ref{fig:abmseq}. The points at which both values go to unity indicate the positions of the word $ab$, $abab$, $ababab$. For all inputs tested, the fidelity between the final state and the final state expected if the word is in $L_{ab}$ is smaller than the Jaro distance between those words.

%
\ 
%


\section{Quantum Inputs}

The inputs used so far have all been classical, represented by a quantum superposition state. However, the way these walks have been set up allows us to give a quantum input. Each symbol in the word can be in a superposition of $a$ or $b$, $ x a + y b $ such that $ |x|^2 + |y|^2 = \alpha^2 $. Superpositions of words, for example $abab$ and $bbbb$ can then be created by using the appropriate superposition for each symbol in the word. Where symbols match, the amplitude is allocated to that symbol as for the classical encoding.  Where symbols do not match, the amplitude is distributed between the $a$ and $b$ states accordingly. 

\begin{center}
\begin{figure}[t]
\centering \includegraphics[scale = 0.4]{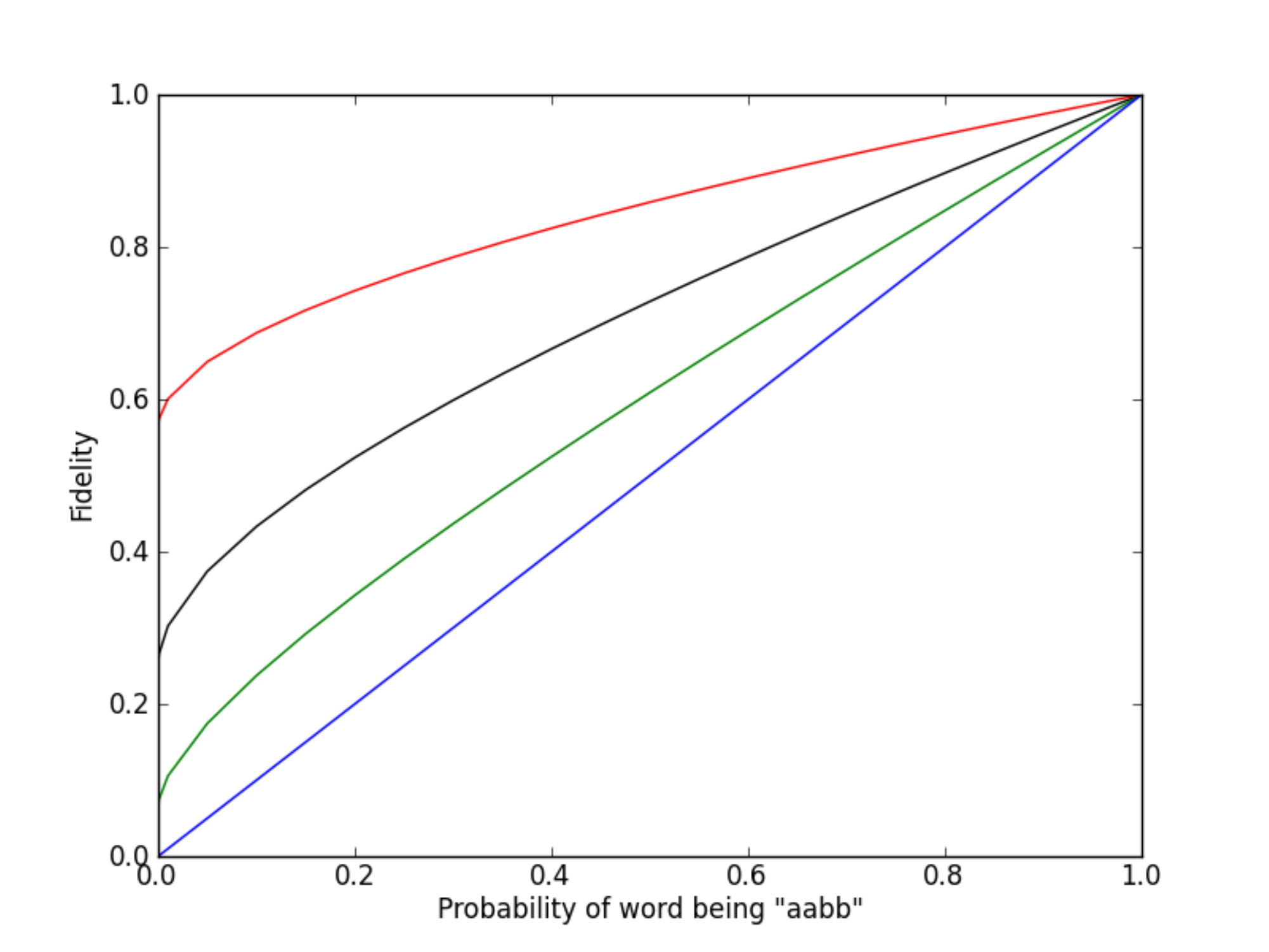}
\caption{Fidelity of final state to accepting state for quantum inputs in a superposition of $aabb$ and: the string with no matching characters, $bbaa$ (blue); the strings with one matching character, $abaa$, $baaa$, $bbab$ and $bbba$ (green); the strings with two matching characters $aaaa$, $abab$, $abba$, $baab$, $baba$ and $bbbb$ (black); the strings with three matching characters $aaab$, $aaba$, $abbb$ and $babb$ (red)} 
\label{fig:qinput}
\end{figure}
\end{center}

As a preliminary investigation into using quantum inputs we tested the effects of using a quantum input on the acceptance probability for the walks accepting $L_{eq}$ using the spatially distributed input. We tested a range of superpositions from the word being entirely $aabb$ to the word being entirely another string, for every string of length four. The effect of the quantum input on the fidelity fell into four distinct cases, as can be seen in Figure \ref{fig:qinput}. These cases correspond to the number of characters a string of length four can differ by, with the probability of being able to accurately determine which input was used diminishing as the number of symbols which match exactly between the two strings increases. 

Using a quantum input frames the question of language acceptance in terms of quantum state discrimination. If we have one of two states $| \psi \rangle $ and $| \phi \rangle $, one of which encodes a word in the language accepted by the walk, and the other does not, then we can gain information about which state we are likely to have had by whether or not it is accepted by the walk. Another setup for using discrete time quantum walks to perform state discrimination by measuring at specific positions is outlined in \cite{QWPOVM}.

\section{Conclusion}


We have presented a novel application of discrete time quantum walks and shown that they can recognise some formal languages using limited spatial resources and numbers of operations. These walks have been tested for numerous inputs and the fidelities between the final state and the state which would be attained if the input was of the required form were calculated. As the fidelities tend to be at a few specific values between 0 and 1, rather than over the entire range, it would be interesting to classify which types of strings give rise to which values for the fidelity. This could aid finding walks to accept further languages. 

In addition to further analysis of the walks described here, there is much future work to be done. It might be nice, for example, to find general ways of specifying graphs accepting arbitrary regular expressions, or even arbitrary formal languages. Or one could restrict attention to specific, appropriately defined subclasses of regular expressions to relate the work to other aspects of logic. In order to compare the walks with other standard models of computation in terms of language acceptance, relations between the efficiency measures used here and other standard efficiency measures such as the minimal number of computational states, or tape squares traversed, must be found.

The two approaches discussed in this paper may not be the only ways to specify quantum walks such that they can be interpreted as accepting formal languages. It is not yet clear what the most fruitful approach will be. A better understanding of the relative merits of using spatially versus sequentially distributed inputs will be informative, and gaining insights into their limitations may suggest further ways of specifying walks to recognise languages. 

We have briefly described how the constructions presented here allow quantum inputs. This new form of input raises many questions. As well as finding out more detail about the state discrimination performed by the walks presented here, it would be interesting to gain a better understanding of the inputs themselves, and how they might form quantum languages.







\newpage
\bibliographystyle{unsrt}
{\footnotesize \bibliography{biblio}}

\end{document}